\documentstyle[12pt,aps]{revtex}
\begin{document}

\begin{titlepage}
\thispagestyle{empty}
\begin{center}
\vspace*{1cm}
{\bf  \Large Asymmetric exclusion model for mixed ionic conductors}\\[13mm]

{\Large {\sc
Sven Sandow\footnotemark[1] \vspace*{2mm} \\
 Steffen Trimper\footnotemark[2] \vspace*{4mm}\\David Mukamel}\footnotemark[1]
} \\[5mm]

\begin{minipage}[t]{13cm}
 \begin{center}
 {\small\sl
\footnotemark[1] Department of Physics of Complex Systems,\\
 The  Weizmann Institute of Science, \\ Rehovot
76100, Israel \\[2mm]
\footnotemark[2] Fachbereich Physik,\\
 Martin-Luther-Universit\"at Halle-Wittenberg\\
 D-06108 Halle, Germany}
 \end{center}
\end{minipage} \vspace{10mm}
 \end{center}
{\small The ionic conductivity of
mixed alkali glasses exhibits a deep minimum as a function of the relative
concentrations of the two alkali ions. To study this behaviour we consider a
simple one-dimensional model for asymmetric diffusion of two kinds of
particles.
Different particles are assumed to repulse each other. We consider
two versions of the model: with or without overtaking of particles.  For the
case of perfect repulsion we find exact expressions for the stationary current.
The model with weaker repulsion is studied by means of numerical simulations.
The stationary current as a function of the ratio of particle concentrations is
found to exhibit a minimum, related to correlations existing in this system.}
\\
\vspace{5mm}
\end{titlepage}

\newpage
\section{Introduction}

Interesting physical effects have been  observed in studying the behaviour of
glassy ionic conductors. The conductivity of mixed alkali systems exhibits a
strong  dependence on the ratio of the alkali concentrations with a deep
minimum at a ratio close to $1$ \cite{da}. In some experiments the minimum
conductivity is  a few orders of magnitude  smaller than the conductivity of
the pure system.  An interesting model explaining  the anomalous conductivity
has  recently been introduced in \cite{mabuin} (see also \cite{buinmang},
\cite{bumain}).  The model is based on the assumption that cations in glasses
create and maintain their own local environment which, due to memory effects,
produce a strong dependence of the conductivity on the relative concentrations
of the two ions. Numerical studies of this model have been presented.

Problems of this kind can be understood within the framework of lattice gas
models. Recently an extensive effort has been invested in studying particles
hopping in a preferred direction with stochastic dynamics and hard core
interactions. These are simple examples of non-equilibrium macroscopic systems
\cite{sp}, \cite{kalesp} which exhibit very interesting collective phenomena
such as phase transitions. These models have been used to study hopping
conductivity or diffusion in narrow pores\cite{kalesp},\cite{ri}. Furthermore,
they are related to  growth processes \cite{mera}-\cite{kamu}.  Some
exact results for particular models are known \cite{li}-\cite{dejale}.

The aim of this paper is to discuss a lattice gas model for asymmetric
particle hopping which is related to the behaviour of mixed ionic conductors.
The model describes the dynamics of two kinds of particles moving  in the same
direction under the influence of a driving field. Here, we do not consider
site memory effects but rather assume that there exists a repulsive
interaction between the two types of ions.  Our model, studied in $d=1$
dimensions, is simpler than the one introduced in \cite{mabuin}.

We consider two versions of the model. In the first version the two kinds of
particles jump independently to the right along a one-dimensional lattice
without overtaking each other.  In the second version different kinds of
particles are allowed to overtake. This may incorporate the effect of the
higher space dimension into the one-dimensional model.

\section{The Model} \setcounter{equation}{0} To model the dynamics of a system
with mixed charge carriers we consider a one-dimensional lattice of length $L$.
Each site can take one of three states: it may be occupied by a particle of
type $a$ or by a particle of type $b$, or it may be vacant (occupied by a hole
$e$). We assume that both kinds of particles  contribute to the conductivity of
the  system. The $a$- and $b$- particles may be identified with different
alkali ions. The
particles are assumed to undergo an asymmetric exclusion dynamics. The
asymmetry is caused by a strong electric field. Throughout this paper we assume
periodic boundary conditions.

The total number of $a$- ($b$-) particles is assumed to be equal to $\rho_a L$
( $\rho_b L$ ). We define $r$ as the ratio of concentration of $a$-particles to
the sum of both particle concentration.
\renewcommand{\theequation}{\arabic{section}.\arabic{equation}}
\begin{equation} \label{B22}
r=\rho_a/\rho\;\;\;\mbox{with}\;\;\;\rho=\rho_a+\rho_b
\;\;.\end{equation}

To be specific we consider stochastic asymmetric exclusion dynamics with
interaction between the $a$- and $b$-particles. Two versions of the model are
studied. In the first version no overtaking of $a$- and $b$-particles is
allowed. A particle can move to its right if the site on its right is vacant.
The probability of making the step during the time interval $dt$ depends on
whether an $ab$-bond is broken ($\alpha dt$), created ($\beta dt$), neither
broken nor created $(1dt)$ or one bond is broken and another one created
($\alpha \beta dt$). The interaction between the two types of particles
is thus introduced by the two parameters $\alpha$ and $\beta$. The possible
steps defining the dynamics are given by the following processes:
\begin{eqnarray}
\label{B01}
xaex~~~&\Longrightarrow&~~~xeax~~~\mbox{with rate}\quad 1 \\
\label{B02}
baex~~~&\Longrightarrow&~~~beax~~~\mbox{with rate}\quad \alpha \\
\label{B03}
xaeb~~~&\Longrightarrow&~~~xeab~~~~\mbox{with rate}\quad \beta \\
\label{B04}
baeb~~~&\Longrightarrow&~~~beab~~~~\mbox{with rate}\quad \alpha \beta\\
\label{B05}
ybey~~~&\Longrightarrow&~~~yeby~~~\mbox{with rate}\quad 1 \\
\label{B06}
abey~~~&\Longrightarrow&~~~aeby~~~\mbox{with rate}\quad \alpha \\
\label{B07}
ybea~~~&\Longrightarrow&~~~yeba~~~~\mbox{wth rate}\quad \beta \\
\label{B08}
abea~~~&\Longrightarrow&~~~aeba~~~~\mbox{with rate}\quad \alpha \beta
\end{eqnarray}
No other steps are allowed. In these equations the occupation of the four
sites $i-1,i,i+1,i+2$ is given and the dynamical step takes place between
sites $i$ and $i+1$. We denote by $x$  a site which is  occupied by either an
$a$-particle or by a hole $e$, and $y$ represents a site which is either
occupied by a $b$-particle or by a hole $e$. The interaction between the two
kinds of particles is attractive for $\alpha<1\;,\;\beta>1$ and  is
repulsive for $\alpha>1\;,\;\beta<1$. The other regions of the
$\alpha,\beta$-plane describe dynamics with competing interactions.

One can look at $d>1$ dimensional systems as described by coupled chains which
are directed along the driving field. The coupling between the chains allows
for hopping of particles from one chain to another. As a result overtaking of
$a$- and $b$-particles in a single chain can take place by hopping via
neighbouring chains. One may therefore model certain aspects of higher
dimensional systems  by studying a one-dimensional model with overtaking. The
rate $\gamma$ of overtaking events is expected to be low. We thus consider a
second version of the model where the following steps, involving overtaking,
are permissible in addition to those given by Eqs. (\ref{B01})-(\ref{B08}):
\begin{eqnarray}
\label{B1}
ea(e)be \;\; &\Longrightarrow&\;\;eb(e)ae \;\;\mbox{with rate} \;\gamma\\
\label{B1a}
aa(e)be \;\; &\Longrightarrow&\;\;ab(e)ae \;\;\mbox{with rate} \;\beta\gamma\\
\label{B1b}
ba(e)be \;\; &\Longrightarrow&\;\;bb(e)ae \;\;\mbox{with rate} \;\alpha\gamma\\
\label{B1d}
ea(e)ba \;\; &\Longrightarrow&\;\;eb(e)aa \;\;\mbox{with rate} \;\alpha\gamma\\
\label{B1e}
aa(e)ba \;\; &\Longrightarrow&\;\;ab(e)aa \;\;\mbox{with rate} \;\alpha \beta
\gamma\\
\label{B1f}
ba(e)ba \;\; &\Longrightarrow&\;\;bb(e)aa \;\;\mbox{with rate}
\;{\alpha}^2\gamma\\
\label{B1g}
ea(e)bb \;\; &\Longrightarrow&\;\;eb(e)ab \;\;\mbox{with rate} \;\beta\gamma\\
\label{B1h}
aa(e)bb \;\; &\Longrightarrow&\;\;ab(e)ab \;\;\mbox{with rate} \;{\beta}^2
\gamma\\
\label{B1c}
ba(e)bb \;\; &\Longrightarrow&\;\;bb(e)ab \;\;\mbox{with rate} \;\alpha \beta
\gamma
\;\;.
\end{eqnarray}
Another set of allowed processes are obtained from (\ref{B1})-(\ref{B1c}) by
 interchanging $a$ and $b$.
 The symbol $(e)$ means, there can  be a hole between the $a$- and the $b$-
particle, which stays at the same position during the process.
 Note that  the processes with a hole between the particles
involve a next-nearest-neighbour interchange.

\section{Mean field approximation} \setcounter{equation}{0}

 In this section we discuss the stationary current in mean field
approximation (MFA). The mean field approximation gives the
same stationary
current for both versions of the model.
 The stationary current of $a$ particles in MFA can be written
\begin{equation}
j_{a} = \rho_{a} ( 1 - \rho_{b} )^{2} ( 1 - \rho ) + ( \alpha + \beta )
\rho_{a} \rho_{b} ( 1 - \rho_{b} ) ( 1 - \rho )  + \alpha \beta \rho_{a}
\rho_{b}^{2} (1 - \rho)
\end{equation}
The expression for the $b$-particle current is similar.
In terms of the ratio $r$ introduced in Eq. (\ref{B22}) we get for the total
current $j = j_{a} + j_{b}$
\begin{equation}
j = \rho (1 -\rho) +  r(1-r)\rho^{2} (1-\rho)
\lbrace 2(\alpha+\beta-2)  +(\alpha \beta-\alpha-\beta+1)\rho \rbrace
\label{strom}
\end{equation}
The current exhibits  a minimum in the conductivity if the relation
\begin{equation}
2(\alpha+\beta-2)  +(\alpha \beta-\alpha-\beta+1)\rho<0
\end{equation}
is satisfied.\\

Typical currents $j$ as a function of $r$ for different values of $\rho$ are
given in Fig.1. Although the current exhibits a  minimum at $r=1/2$ (which is
most pronounced for $\alpha=\beta=0$), its value at the minimum is rather close
to the current of the pure system ( $j(1)=j(0)=\rho (1-\rho)$ ), unlike the
experimental
results. In the next section we calculate the exact current for $\beta=0$ and
demonstrate that the current at $r=1/2$ can become much lower than its mean
field value.

\section{Exact results and simulations}
\setcounter{equation}{0}
\subsection{The model without overtaking}

 We now discuss the first version of the model
(rules (\ref{B01}) - (\ref{B08})). Let us  view the ring as a chain
with sites $1,...,L$ and  denote a configuration by  $\underline n=\lbrace n_i
\rbrace$ where  $n_i=a\;,\;b \;\mbox{or}\; e$ if site $i$ is occupied by either
an
$a$-particle, $b$-particle or a hole, respectively.
  Furthermore we define a frame $\underline F$ as the sequence of $a$- and
$b$-symbols which is obtained from
a configuration $\underline n$  by removing all
$e$-symbols. For example the frame  associated with the
configuration $abebbeeaab...$ is $abbbaab...\;$.

The dynamics of the first version of
 the model allow for jumps of $a$- or
$b$-particles into holes but do not allow for exchange of position of two
particles. It may be viewed as the dynamics of holes moving to the left on a
lattice certain sites of which are occupied by particles. Consequently, the
frame $\underline F$  changes in time only by means of cyclic permutations. The
latter ones are possible because of the periodic boundary conditions assumed
throughout this work.
The dynamics are thus nonergodic, and the final
state depends on the initial frame $\underline
F^{(i)}$ defined by the initial configuration $\underline
n^{(i)}$.
\\[0.5cm]

{\em The case $\alpha=\beta=1$ .} Although the experimentally observed anomaly
is not expected here, a short discussion of this case seems  instructive.
There is neither repulsion nor attraction between the  $a$- and $b$-particles
for this choice of the parameters $\alpha$ and $\beta$.   The stationary
probability distribution $p_{st}(\underline n)$ for this process is similar to
the one for exclusive
diffusion of one kind of particle which is well known \cite{spi}. It assigns
the
same probability to all
configurations with  the initial frame $\underline
F^{(i)}$ or its cyclic permutations. The
stationarity of this distribution can be  easily checked by counting the
number of incoming and the number of outgoing states for a given configuration.
(An incoming state is a configuration which can change to the
configuration of interest during one hopping event. Outgoing states are created
by a single event taking place in the given configuration.)
Note that the
form of the stationary distribution depends strongly on the boundary
conditions. The above discussion is correct only for periodic boundary
conditions.

Obviously, since there is no interaction between the $a$- and $b$-particles
beyond the hard core term, the current $j$ does not depend
on the
ratio $r$. As in the case of one kind of particles it is exactly the same as
calculated in MFA:
\begin{equation} \label{A0}
j(r,\rho)=\rho(1-\rho)
\;\;\;. \end{equation}\\[0.5cm]

{\em The case $\beta=0$ ($\alpha \ne 0)$.} This is a nontrivial but exactly
soluble case, in which perfect repulsion between the $a$- and $b$-particles
takes place.
 Since $\beta=0$, no $ab$-bonds (including  $ba$-bonds) are created. But since
$\alpha \ne 0$ $ab$-bonds may be broken.
 Hence the system runs into
configurations for which the number of $ab$-bonds is minimal.
The stationary state is thus characterized by the number $\rho_{bo}L$ of still
existing
$ab$-bonds.

To study the stationary distribution in detail let us first define the quantity
$\rho_{ab}^{(i)}$ as
$\rho_{ab}^{(i)}=n_{ab}^{(i)}/L$ where $n_{ab}^{(i)}$ is the number of
$ab$-bonds in the initial frame $\underline F^{(i)}$. (Note that
due to the periodic boundary conditions we have to include in $n_{ab}^{(i)}$
bonds which may exist between site $L$ and site $1$.) The number of  $ab$-bonds
in the frame does not change in time for the only allowed changes of the frame
are cyclic permutations. A configuration, however,  has generally less
$ab$-bonds than the frame, since holes may be located between the particles of
such a bond. In order to allow for a configuration without any $ab$-bond the
system needs at least as many holes as number of $ab$-bonds in the frame.
 Hence, if
 $1-\rho \le \rho_{ab}^{(i)}$, the system runs into a configuration with a
 density
$\rho_{bo}=\rho_{ab}^{(i)}-1+\rho$ of $ab$-bonds and all holes are
stuck between  $a$- and  $b$-particles. Since any change of configuration
takes place by means of a hopping hole, the configuration  the system runs into
does not change in time  and has a vanishing current.
On the other hand, if
$1-\rho > \rho_{ab}^{(i)}$, the
number of holes exceeds the  number of $ab$-bonds  in the frame.
 The system evolves
into configurations with no $ab$-bonds, i.e., where $a$- and $b$-particles are
separated by at least one hole.  The number of holes free to hop is given by
$n_f=\rho_fL$ with
\begin{equation}\label{A1}
\rho_f=1-\rho -\rho_{ab}^{(i)}
\end{equation}
It turns out that for $1-\rho > \rho_{ab}^{(i)}$ the distribution assigning the
same probability to all
configurations with no $ab$-bonds and with  the frame
 $ \underline F^{(i)}$ or its cyclic permutations is stationary. This
can be seen  by noting that the number of incoming states for any such
configuration
$ \underline n$ is equal to the number of states to which the configuration can
evolve.

These simple considerations enable one to
derive a general expression for the current $j(r,\rho)$. Let us consider the
case of nonvanishing current ($1-\rho > \rho_{ab}^{(i)}$). A jump between two
neighbouring  sites, say site $1$ and site $2$, occurs
with  rate $1$ if the left site is occupied by a particle and the right one by
a free hole (a hole which is not stuck between an $a$- and an
$b$-particle). Therefore $j(r,\rho)$ equals the probability of finding $ae_f$
 or $be_f$
at sites $1\; 2$, where $e_f$ denotes a free hole. It reads
\begin{equation}
\label{j1}
j(r,\rho)=prob\;(n_1=a \;\mbox{or}\;n_1=b)\;
prob\;(n_2=e_f|n_1=a \;\mbox{or}\;n_1=b)
\;\;.\end{equation}
Here, the second term in the right hand side of this equation denotes the
conditional probability of finding a free hole at site $2$ given that site $1$
is occupied by a particle.  Due to translational invariance one has
$prob\;(n_1=a \;\mbox{or}\;n_1=b)=\rho$.
The probability $prob\;(n_2=e_f|n_1=a \;\mbox{or}\;n_1=b)$ is given by the
ratio
of the number  of configurations which have a free hole at site $2$ and a
particle at site $1$ to the total number of
 configurations  with a particle at site $1$.
 The configurations contributing to these numbers may have different frames.
However, all these frames can be obtained
from $\underline F^{(i)}$
by cyclic permutation. Thus they all have the same number of $ab$-bonds, and we
may restrict ourselves to configurations with the frame $\underline F^{(i)}$.
The above probability is given by
  $prob\;(n_2=e_f|n_1=a \;\mbox{or}\;n_1=b)= X /Y$,
where $Y$ is the number of
 configurations  with the frame $\underline F^{(i)}$ and a particle at site
$1$,
and
$ X$ is the number of those configurations  which in addition have a free hole
at site $2$.
All  configurations to be counted
may be constructed
  by first inserting  a hole
 in between any pair of $ab$-particles in the frame $\underline F^{(i)}$.
One then has to distribute the remaining $n_f=\rho_f L$ free holes in between
the particles in a way that site $1$ is occupied by a particle.
Thus $Y$ is equal to the number of ways of distributing $n_f$
indistinguishable
holes
in $n=n_a+n_b$ states. It is given by
$Y=\renewcommand{\arraystretch}{0.8}
\mbox{$\left(\begin{array}{@{}c@{}}{\scriptstyle n+n_f-1}
\\{\scriptstyle n_f}\end{array}\right)$}\renewcommand{\arraystretch}{1}$.
Out of these configurations
$X=Y-
\renewcommand{\arraystretch}{0.8}
\mbox{$\left(\begin{array}{@{}c@{}}{\scriptstyle n+n_f-2}
\\{\scriptstyle n_f}\end{array}\right)$}\renewcommand{\arraystretch}{1}$
have a free hole at site $2$.
The second term on the right hand side of this equation gives the number of
configurations with no free hole at site $2$.
The current $j(r,\rho)$ for
$1-\rho > \rho_{ab}^{(i)}$  is therefore given by:
$j(r,\rho)=\rho \;\lbrace\;
\renewcommand{\arraystretch}{0.8}
\mbox{$\left(\begin{array}{@{}c@{}}{\scriptstyle n+n_f-1}
\\{\scriptstyle n_f}\end{array}\right)$}\renewcommand{\arraystretch}{1}
-\renewcommand{\arraystretch}{0.8}
\mbox{$\left(\begin{array}{@{}c@{}}{\scriptstyle n+n_f-2}
\\{\scriptstyle
n_f}\end{array}\right)$}\renewcommand{\arraystretch}{1}\;\rbrace
\renewcommand{\arraystretch}{0.8}
\mbox{$\left(\begin{array}{@{}c@{}}{\scriptstyle n+n_f-1}
\\{\scriptstyle n_f}\end{array}\right)$}\renewcommand{\arraystretch}{1}^{-1}
\;\;$.
Simplifying the binomials results in the following  expression for the current:
\begin{eqnarray} \label{A2}
j(r,\rho)=\left\{
\begin{array}{ll}
\frac{\rho \rho_f}{\rho+\rho_f-L^{-1}} & \mbox{for}\;\;\; 1-\rho >
\rho_{ab}^{(i)}  \\
0 & \mbox{otherwise}
\end{array}
\right.
\end{eqnarray}
The density of free holes, $\rho_f$, is a function of  $\rho$
and    of  $\rho_{ab}^{(i)}$ , as expressed in (\ref{A1}). In the thermodynamic
limit the $L^{-1}$-term in the expression for the current vanishes.

 Consider now the case of random initial conditions in which the initial
configuration $\underline n^{(i)}$ is created by uniformly distributing $\rho_a
L$ $a$-particles and $\rho_bL$ $b$-particles on a lattice of length $L$. The
average number of $ab$-bonds in the frame $\underline F^{(i)}$ corresponding to
this initial
condition is $2r(1-r)\rho L$. For  $L \rightarrow \infty$ the relative
fluctuations of this quantity vanish. Consequently, we find for any initial
configuration with a- and b-particles uniformly scattered on a large lattice:
\begin{equation}\label{A2a}
\rho_{ab}^{(i)}=2r(1-r)\rho
\;\;\;.
\end{equation}
 Combining this with Eq.
(\ref{A1}) and (\ref{A2}) we obtain:
\begin{eqnarray} \label{A3}
j(r,\rho)=\left\{
\begin{array}{ll}
\rho \frac{1-\rho[1+2r(1-r)] }{1-2r(1-r)}& \mbox{for}\;\;\;
\rho< \rho_c(r)  \\
0 & \mbox{otherwise}
\end{array}
\right.
\end{eqnarray}
where
\begin{equation} \label{A3a}
\rho_c(r)=1/[1+2r(1-r)]
\end{equation}
is the critical density. The expressions for the current $j$ and the critical
 density $\rho_c$ are independent of $\alpha$ (as long as $\alpha \ne 0$).
 Fig.1 shows the current as a function of $r$ for
different densities $\rho$. For densities $\rho<\rho_c(r=1/2)=2/3$ the current
is nonzero for any value r.

This result is rather different from the mean field current (\ref{strom}).
 While the latter  is a function of $\alpha$ the exact expression for the
current $j$ is independent of $\alpha$.
Fig. 1 compares the exact result with the mean field current (\ref{strom})
for $\alpha=0$.
This value is chosen since it corresponds to the most pronounced minimum of the
conductivity in mean field approximation.
The discrepancy between mean field and exact results
indicates that the deep minimum is related to correlations. \\[0.5cm]

{\em  The case $\beta > 0$ ($\alpha \ne 0$) .}
For nonvanishing $\beta$ the stationary distribution is more difficult to
calculate and one has to resort to numerical simulations of the model. However,
 a few properties of the $j(r,\rho)$-function can be
seen easily. For $r=0$ and $r=1$ the current is equal to the mean field current
given in
Eq. (\ref{A0}) since in both cases only one kind of particles performs
exclusive diffusion. The stationary solution for that process is known to be
uncorrelated \cite{sp}.
 Furthermore the $j(r)$-function is symmetric with respect to reflection about
 the
$r=1/2$-line because the model is defined in way that a- and b-particles play
the same role.

The stationary current for $\beta > 0$ does not vanish for high densities, as
in the case $\beta=0$ since there are always possible hopping events.
Consequently, we  expect a qualitatively different  behaviour for $\rho>\rho_c$
as compared to the  $\beta=0$-case. Replacing $\beta=0$ by a finite $\beta$
 the current is increased for any value of $\rho$ and $r$.


Computer simulations of the process were performed by letting particles hop
stochastically on a lattice of $L=1000$ sites. Averages are calculated as time
averages for a particular realization as is done in experiments. Besides, time
averages and ensemble averages coincide if the stationary probability
distribution is chosen properly, i.e. in the subspace of the phase space which
is actually reached by the system.

We have carried out simulations for the case $0<\beta << 1$ and
$\rho=0.7>\rho_c$ which seems to reflect the experimental results for the
current \cite{da} quite well. The parameter $\alpha$ is chosen to be $1$ but it
is supposed to play a minor role as long as $\alpha \ne 0$. Results are shown
in
Fig.2. Tuning $\beta$,
which depends on the temperature via $\beta=exp(-\delta E/k_B T)$, any ratio
$j(r=0)/j(r=0.5)$ can be obtained for $\rho > \rho_c$. Here $\delta E$ is the
energy barrier one has to overcome by creating an $ab$-bond. The model is
therefore
capable of exhibiting the deep minimum of the conductivity which is observed
experimentally.\\[0.5cm]

\subsection{The model with overtaking}

 We next study the second version of the model (rules (\ref{B01})-(\ref{B1c}))
where $\gamma \ne 0$.  Since overtaking is allowed in these dynamics, the frame
undergoes noncyclic permutations of particles unlike in the $\gamma=0$ case.

Let us discuss the case of perfect repulsion ($\beta=0$) in detail. By simple
state-counting it can be shown that a distribution assigning the same
probability to any occurring configuration is stationary. But while only
configurations with the initial frame $\underline F^{(i)}$ and its cyclic
permutations are allowed for the first version of the
model, here there is a broader distribution of frames which are reached by the
dynamics.
As in the previous version, for sufficiently  large density $\rho$ the model is
expected to run into a configuration in which all holes are stuck between $a$-
and $b$-particles. This state has a vanishing current. On the other hand, for
densities lower than some critical density $\rho_c(r) $ the number of holes
exceeds the number of $ab$-bonds in the initial frame. The system therefore has
free holes which generate nonvanishing current.

For densities satisfying $\rho\le \rho_c(r)$ rules (\ref{B1})-(\ref{B1c}) allow
for changes in the frame via interchanges of $a$- and $b$-particles. Any frame
 with a number of bonds less than
the number of holes can be created during the dynamics. The stationary
probability distribution $p_{st}$ assigns the same probability to any
configuration without $ab$-bonds.

To calculate the current associated with the  stationary distribution we first
consider the current averaged over all configurations whose frames have
 $n_{ab}=\rho_{ab}L$
$ab$-bonds. As it was shown in Section IV A the  current averaged over all
configurations with $\rho$ and $\rho_{ab}$ is given by
\begin{equation} \label{B2}
j(\rho,\rho_{ab})=\frac{\rho \rho_{f}}{\rho+\rho_f-L^{-1}}
\end{equation}
where
\begin{equation} \label{B2a}
\rho_{f}=1-\rho-\rho_{ab}
\end{equation}
is the density of the free holes. The steady state current is thus obtained by
averaging $j(\rho,\rho_{ab})$ over all possible $\rho_{ab}$ or $n_{ab}$. Since
all allowed configurations have the same weight
one has to find the number of configurations corresponding to $n_{ab}$. To this
end we consider first the number of frames
associated  $n_{ab}$.
We then calculate the number of ways of distributing $n_f=\rho_f L$ free holes
on a given frame, and obtain the probability of having a frame with
$n_{ab}$ $ab$-bonds.
Note that, as explained at the beginning of Section IV A, a frame is defined by
taking a configuration $\underline n=\{n_1,...,n_L\}$ and removing all holes.
This defines  a sequence of $A$ intervals of $a$-particles alternating with $B$
intervals of $b$-particles. The first and the last interval may either be of
the
same or of different type.
If the intervals at both ends of the lattice are the same, say $a$, type then
$A=n_{ab}/2+1$ and $B=n_{ab}/2$. Similarly if both end intervals are of
$b$-type then $A=n_{ab}/2$ and $B=n_{ab}/2+1$. On the other hand if the two end
intervals are of different types then $A=B=n_{ab}/2$. The number of ways of
arranging $n_a=r \rho L$ $a$-particles in   $A$ groups is given by
$\renewcommand{\arraystretch}{0.8}
\mbox{$\left(\begin{array}{@{}c@{}}{\scriptstyle n_a-1}
\\{\scriptstyle A-1}\end{array}\right)$}\renewcommand{\arraystretch}{1}$,
and similarly the number of possibilities of arranging the
$n_b=(1-r) \rho L$ $b$-particles in $B$ groups is
$\renewcommand{\arraystretch}{0.8}
\mbox{$\left(\begin{array}{@{}c@{}}{\scriptstyle n_b-1}
\\{\scriptstyle B-1}\end{array}\right)$}\renewcommand{\arraystretch}{1}$.
Therefore in each of the above cases the number of possible frames is
$\renewcommand{\arraystretch}{0.8}
\mbox{$\left(\begin{array}{@{}c@{}}{\scriptstyle n_a-1}
\\{\scriptstyle A-1}\end{array}\right)$}\renewcommand{\arraystretch}{1}
\renewcommand{\arraystretch}{0.8}
\mbox{$\left(\begin{array}{@{}c@{}}{\scriptstyle n_b-1}
\\{\scriptstyle B-1}\end{array}\right)$}\renewcommand{\arraystretch}{1}$.
We now have to find the number of ways of
distributing $n_f=\rho_f L$ free holes in each frame. In the case where both
ends of the frame are of the same type this number is given by
$\renewcommand{\arraystretch}{0.8}
\mbox{$\left(\begin{array}{@{}c@{}}{\scriptstyle L-n_{ab}}
\\{\scriptstyle n_f }\end{array}\right)$}\renewcommand{\arraystretch}{1}$
 while in the case where the two ends are different it is
given by $[\;
\renewcommand{\arraystretch}{0.8}
\mbox{$\left(\begin{array}{@{}c@{}}{\scriptstyle L-n_{ab}}
\\{\scriptstyle n_f }\end{array}\right)$}\renewcommand{\arraystretch}{1}
+
\renewcommand{\arraystretch}{0.8}
\mbox{$\left(\begin{array}{@{}c@{}}{\scriptstyle L-n_{ab}-1}
\\{\scriptstyle n_f }\end{array}\right)$}\renewcommand{\arraystretch}{1}\;]$.
Thus the
statistical weight associated with $n_{ab}$ takes the form
\begin{eqnarray}
f(n_{ab})&=&[\;
\renewcommand{\arraystretch}{0.8}
\mbox{$\left(\begin{array}{@{}c@{}}{\scriptstyle n_a-1 }
\\{\scriptstyle n_{ab}/2 }\end{array}\right)$}\renewcommand{\arraystretch}{1}
\renewcommand{\arraystretch}{0.8}
\mbox{$\left(\begin{array}{@{}c@{}}{\scriptstyle n_b-1 }
\\{\scriptstyle n_{ab}/2-1 }\end{array}\right)$}\renewcommand{\arraystretch}{1}
+
\renewcommand{\arraystretch}{0.8}
\mbox{$\left(\begin{array}{@{}c@{}}{\scriptstyle n_a-1 }
\\{\scriptstyle n_{ab}/2-1 }\end{array}\right)$}\renewcommand{\arraystretch}{1}
\renewcommand{\arraystretch}{0.8}
\mbox{$\left(\begin{array}{@{}c@{}}{\scriptstyle n_b-1 }
\\{\scriptstyle n_{ab}/2 }\end{array}\right)$}\renewcommand{\arraystretch}{1}
\;]\;
\renewcommand{\arraystretch}{0.8}
\mbox{$\left(\begin{array}{@{}c@{}}{\scriptstyle L-n_{ab} }
\\{\scriptstyle n_f }\end{array}\right)$}\renewcommand{\arraystretch}{1}
\nonumber\\
&&+2
\renewcommand{\arraystretch}{0.8}
\mbox{$\left(\begin{array}{@{}c@{}}{\scriptstyle n_a-1 }
\\{\scriptstyle n_{ab}/2-1 }\end{array}\right)$}\renewcommand{\arraystretch}{1}
\renewcommand{\arraystretch}{0.8}
\mbox{$\left(\begin{array}{@{}c@{}}{\scriptstyle n_b-1 }
\\{\scriptstyle n_{ab}/2-1 }\end{array}\right)$}\renewcommand{\arraystretch}{1}
\;[\;
\renewcommand{\arraystretch}{0.8}
\mbox{$\left(\begin{array}{@{}c@{}}{\scriptstyle L-n_{ab} }
\\{\scriptstyle n_f}\end{array}\right)$}\renewcommand{\arraystretch}{1}
+
\renewcommand{\arraystretch}{0.8}
\mbox{$\left(\begin{array}{@{}c@{}}{\scriptstyle L-n_{ab}-1 }
\\{\scriptstyle n_f}\end{array}\right)$}\renewcommand{\arraystretch}{1}
\;]\nonumber\\\label{B2b}
&=& \frac{2L}{n_{ab}}
\renewcommand{\arraystretch}{0.8}
\mbox{$\left(\begin{array}{@{}c@{}}{\scriptstyle L-n_{ab}-1 }
\\{\scriptstyle (1-\rho)L-n_{ab} }\end{array}\right)$}
\renewcommand{\arraystretch}{1}
\renewcommand{\arraystretch}{0.8}
\mbox{$\left(\begin{array}{@{}c@{}}{\scriptstyle r\rho L-1 }
\\{\scriptstyle n_{ab}/2-1 }\end{array}\right)$}
\renewcommand{\arraystretch}{1}
\renewcommand{\arraystretch}{0.8}
\mbox{$\left(\begin{array}{@{}c@{}}{\scriptstyle (1-r)\rho L-1 }
\\{\scriptstyle n_{ab}/2-1 }\end{array}\right)$}
\renewcommand{\arraystretch}{1}
\end{eqnarray}

Using Eqs. (\ref{B2})-(\ref{B2b}) we find for the stationary current:
\begin{eqnarray} \label{B6}
j(r,\rho)=\left\{
\begin{array}{ll}
\rho Z^{-1}\;\sum_{k=0}^{\infty}\frac{1-\rho-2 k L^{-1}}{1-(2 k+1) L^{-1}}
f(2k)
& \mbox{for}\;\;\;  \rho <\rho_c(r)  \\
0 & \mbox{otherwise}
\end{array}
\right.
\;\;\;\end{eqnarray}
where $Z=\sum_{k=0}^{\infty} f(2k)$ is a normalization constant.

 Equations (\ref{B2b})-(\ref{B6}) give the stationary current for a lattice of
arbitrary length $L$. In the thermodynamic limit  the weight function
$f(n_{ab})$ is sharply peaked around $n_{ab}={\rho_{ab}}^*L$ where
${\rho_{ab}}^*$ is a
solution of the following equation:
\begin{eqnarray}\label{B7}
0&=&
r(1-r)\rho(1-\rho)^2-
\frac{1}{2}(1-\rho)[1-\rho+4  r(1-r)\rho]\;{\rho_{ab}}^*
\nonumber\\
&&+[\frac{1}{2}-\frac{3}{4}\rho+r(1-r)\rho]\;({{\rho_{ab}}^*})^2
\;\;\;.
\end{eqnarray}
This yields ${\rho_{ab}}^*$ is a function of $\rho$ and $r(1-r)$ which has its
maximum at
$r=1/2$ for any given value of $\rho$.  For example , solving
Eq.(\ref{B7}) for $r=1/2$ we find
\begin{equation} \label{B7a}
{\rho_{ab}}^*(r=\frac{1}{2},\rho)=
\frac{1}{2}-\frac{1}{2}\sqrt{1-2\rho(1-\rho)}
\;\;\;.
\end{equation}

Hence, for an infinite lattice the number of bonds in the frame is
${\rho_{ab}}^*L$
and the stationary current is
\begin{eqnarray} \label{B8}
j(r,\rho)=\left\{
\begin{array}{ll}
\rho \frac{1-\rho-{\rho_{ab}}^*(r,\rho)}{1-{\rho_{ab}}^*(r,\rho)}  &
\mbox{for}\;\;\;
 \rho<\rho_c(r)  \\
0 & \mbox{otherwise}
\end{array}
\right.
\;\;\;.\end{eqnarray}

Let us now consider the random initial conditions defined in Section IV A
and try to estimate the critical density $\rho_c(r)$ below which the current
of the system is nonzero. Clearly, if the  concentration
$\rho_{ab}^{(i)}=2r(1-r) \rho$ of $ab$-bonds in the initial frame
$\underline F^{(i)}$ satisfies $\rho_{ab}^{(i)}<1-\rho$, the system has free
holes
which generate ergodic dynamics leading to a current $j$ as given by Eqs.
(\ref{B7a}) and (\ref{B8}). This yields a first lower bound for the critical
density
$\rho_c(r)>1/[1+2r(1-r)]$. However, due to the particle exchange mechanism
which exists for $\gamma>0$ the system may exhibit a finite current even for
$\rho_{ab}^{(i)}\gtrsim 1-\rho$. This may easily be seen by considering the
limit $\gamma \ll 1$. In this limit and for time scales shorter than $1/
\gamma$, basically all initially free holes are caught by $ab$-bonds, and thus
stop moving. However for longer time scales, where particle exchange processes
take place, some of the holes which are stuck in $ab$-bonds are released. For
example a sequence $abab$ may evolve into $aabb$ thus reducing the number of
$ab$-bonds by two and creating free holes.  And
 a sequence $aabbaabb$ may in principle evolve into $aaaabbbb$, but this
requires the existence of free holes to begin with. If the only holes in this
sequence are those stuck in the $ab$-bonds, particle exchange processes do  not
take place. However, if in addition there are some free holes, the sequence may
change. If we take into account only changes  which do not require the
existence of free holes the  number of $ab$-bonds in the frame is
reduced and becomes $\overline\rho_{ab}L$ with
\begin{equation} \label{B8a0} \overline
\rho_{ab}=\rho_{ab}^{(i)}-[4r^2(1-r)^2+o(r^3(1-r)^3)] \rho
\;\;.\end{equation}
We thus
expect that for $\overline \rho_{ab}<1-\rho$ the system exhibits a nonvanishing
current. This yields the following lower bound for $\rho_c(r)$:
\begin{equation} \label{B8a}
\rho_c(r) \gtrsim 1/[1+2r(1-r)-4r^2(1-r)^2]+o(r^3(1-r)^3)
\;\;\;.  \end{equation}

The current $j(r,\rho)$ is shown in Fig.3 as a function of $r$ for various
values of $\rho$. In the figure $\rho_c(r)$ is determined by Eq. (\ref{B8a}) to
second order in $r(1-r)$.

The  dynamics of the hopping holes have a time scale of order $1$ whereas the
reordering of the frame has a characteristic time much larger than one since
we consider the case
$\gamma\ll1$. Hence we observe the following scenario: The system runs into the
stationary state of the first version of the model (see Fig. 1) which decays
very slowly to the final state (Fig. 3).

\section{Conclusion}

A simple model describing the transport properties of mixed ionic conductors
has been introduced and analyzed. The model exhibits a minimum conductivity for
equal concentrations of the two species of particles ($r=1/2$), a result which
is compatible with experimental observations in mixed alkali glasses. A mean
field approximation yields the correct qualitative behaviour of the
conductivity, but it fails to explain the very low
conductivity for $r=1/2$. It has been demonstrated by exact solution and
numerical simulations in the strong repulsion limit that the conductivity
corresponding to the model is indeed very small at $r=1/2$ in accordance with
experimental observations.

The model can be extended to study the case in which particles move into both
directions with arbitrary rates. The exact results obtained above for the case
$\beta=0$ are easy to generalize: The probability distribution which assigns
the same probability to all configurations with no bonds is stationary even
when hops in both direction  take place. Moreover, the rates $p$ and $q$ of
hopping to the right and left, respectively, may be taken as time dependent
without changing the stationary state.  The current is then $[p(t)-q(t)]j$,
where $j$ is given by  Eq.(\ref{A2}) or Eq.(\ref{A3}) for $\gamma=0$ and by
Eq.(\ref{B6}) or Eq.(\ref{B8}) when overtaking is included, i.e. for $\gamma
\ne 0$. Obviously, the conductivity does not depend on the frequency if the
driving field is harmonic.\\[1cm]

\begin{center}
{\Large Acknowledgments}
\end{center}
One of us (S. Sandow) gratefully acknowledges financial support by the
 Minerva Foundation.\\[2cm]

\vspace{2cm}

\begin{center}
{\Large Captions to the figures}
\end{center}

{\em Fig.1:}
Stationary current $j$ for the model without overtaking as a function of the
ratio $r$ for $\beta=0$ and
different values of the density $\rho$, solid lines: exact results ($\alpha$
arbitrary), dashed lines: mean field approximation ($\alpha=0$); (1):
$\rho=0.15$; (2):  $\rho=2/3$ ; (3): $\rho=0.8$
\\[1cm]
{\em Fig.2:}
Stationary current $j$ for the model without overtaking as a function of the
ratio $r$ for $\rho=0.7$ , $\alpha=1$ and
different $\beta$`s obtained by simulations (The labels are the values
of $\beta$, and the lines are interpolations.) \\[1cm]
{\em Fig.3:}
Stationary current $j$ for the model with overtaking as a function of the ratio
$r$ for
$\beta=0$, arbitrary $\alpha$ and different values of the density $\rho$, exact
results

\end{document}